\documentclass{sf2a-conf2018}
\usepackage{graphicx}
\usepackage{hyperref}
\usepackage[]{natbib}  
\usepackage{epstopdf}

\def\BibTeX{{\rm B\kern-.05em{\sc i\kern-.025em b}\kern-.08em
    T\kern-.1667em\lower.7ex\hbox{E}\kern-.125emX}}
\bibpunct{(}{)}{;}{a}{}{,}  

\begin{document}

\TitreGlobal{SF2A 2017}


\title{Constraining the environment and habitability of TRAPPIST-1}

\runningtitle{Constraining TRAPPIST-1}

\author{E. Bolmont}\address{AIM, CEA, CNRS, Universit\'e Paris-Saclay, Universit\'e Paris Diderot, Sorbonne Paris Cit\'e, F-91191 Gif-sur-Yvette, France}





\setcounter{page}{237}


\maketitle


\begin{abstract}
The planetary system of TRAPPIST-1, discovered in 2016-2017, is a treasure-trove of information. 
Thanks to a combination of observational techniques, we have estimates of the radii and masses of the seven planets of this very exotic system. 
With three planets within the traditional Habitable Zone limits, it is one of the best constrained system of astrobiological interest.
I will review here the theoretical constraints we can put on this system by trying to reconstruct its history: its atmospheric evolution which depends on the luminosity evolution of the dwarf star, and its tidal dynamical evolution. 
These constraints can then be used as hypotheses to assess the habitability of the outer planets of the system with a Global Climate Model. 
\end{abstract}



\section{Introduction}
\begin{figure}[ht!]
 \centering
 \includegraphics[width=0.5\textwidth,clip]{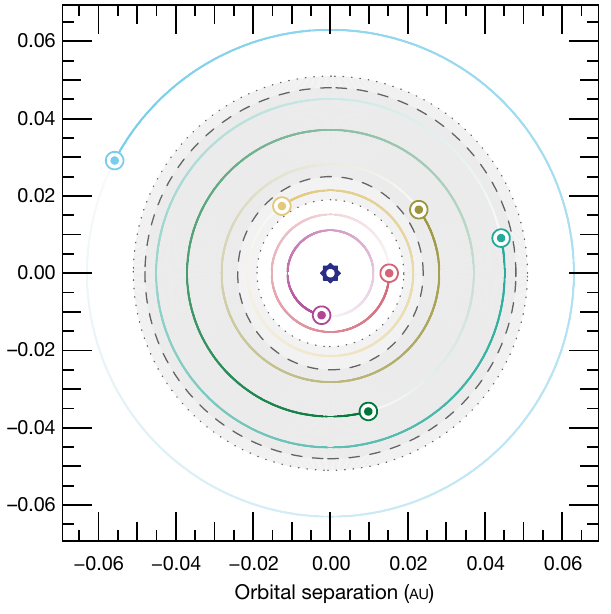}      
  \caption{Representation of the orbits of the TRAPPIST-1 planets (Figure adapted from \citealt{2017Natur.542..456G}).}
  \label{bolmont:fig1}
\end{figure}

The planetary system of TRAPPIST-1 (\citealt{2016Natur.533..221G, 2017Natur.542..456G}; see Figure~\ref{bolmont:fig1}) is a remarquable system for several reasons. 
From the transit measurements from the detection \citep{2016Natur.533..221G,2017Natur.542..456G} and follow-up observations \citep{2017NatAs...1E.129L, 2018MNRAS.475.3577D}, we have good constraints on the orbital periods and radii of the planets and Transit Timing Variations (TTVs) allowed us to have an estimate of their masses and eccentricities \citep{2018A&A...613A..68G}. 
Moreover, the atmosphere of the inner planets of this system, including some Habitable Zone (HZ) planets will be probed by the JWST (with a potential detection of ozone: \citealt{2016MNRAS.461L..92B}, or with a potential identification of the type of the atmosphere: \citealt{2017ApJ...850..121M}). 

However, TRAPPIST-1 is a system very different from our own and it had an entirely different history. 
First, its star is much fainter and redder than the Sun ($L_\star = 10^{-3.28}~L_\odot$). 
This has consequences on the atmosphere of the planet, which is irradiated with light that has a spectral distribution different from the Sun \citep{2005AsBio...5..706S, 2011A&A...529A...8R}.
Moreover, low-mass stars can be very active and the chemical balance of the atmosphere can be modified by flares \citep{2016ApJ...830...77V} and the UV radiations could impact any potential life on the surface \citep[e.g.][]{2016A&A...585A..96T}.
This also means that due to the cooling down of the star with time, the HZ planets were once too hot to sustain surface liquid water. 
During this phase, the water is in gaseous form in the atmosphere and submitted to the high energy radiations that can lead to water loss.
 
Second, the planets are located extremely close-in, within a distance of 0.06~AU. 
This means that star-planet interactions should play a major role in the evolution of the system. 
When the system was young, the stellar tide (the tide raised by the planets in the star) could have driven the orbital migration of the inner planets \citep{2011A&A...535A..94B, 2017haex.bookE..62B}.
The planetary tide (tide raised by the star in a planet) tends to synchronize the planet's rotation and damps the eccentricity of the orbit.

Third, the planets are in an extremely compact system, in a dynamic configuration similar to that of the moons of Jupiter \citep{2017NatAs...1E.129L}.
This means that planet-planet interactions are very strong in the system (which helped to measure the TTVs), which leads to an excitation of the eccentricity of the planets.
This excitation therefore competes with the tidal damping, which can have very concrete consequences: as for Io, the planets can experience an important tidal heating \citep{2017NatAs...1E.129L, 2018A&A...612A..86T}.

Understanding the system of TRAPPIST-1 as it is today requires to investigate all these different aspects. 
Let us first concentrate on the early evolution of TRAPPIST-1, when the star was more luminous and the HZ was located farther out.
Then, we will discuss what we can learn from the system as it is today.
  
\section{Before reaching the HZ}
Before the planets reach the HZ, they experience an evolution of their atmosphere, driven by the radiation of the star, and an evolution of their orbit/rotation, driven by tidal interactions. 

\subsection{Atmospheric evolution}
When the system was young and the star more luminous, the planets in the HZ today were too hot to sustain liquid water. 
The water, in gaseous form in the atmosphere, was submitted to the high energy radiations (E-UV to break the water molecules and X-UV to drive the escape of hydrogen and oxygen atoms) and could escape the planet.
We only consider here planets with water-dominated atmospheres, we only consider the hydrodynamical loss of water and neglect any magnetic interaction which could also drive atmospheric escape. 
This topic was extensively studied for planets around M-dwarfs \citep[e.g.][]{2015AsBio..15..119L} and was applied to the TRAPPIST-1 planets in \citet{2017MNRAS.464.3728B}.
This last study used more recent observations \citep{2010ApJ...709..332B, 2014ApJ...785....9W} than in \citet{2015AsBio..15..119L} and it used an improved energy-limited escape formalism \citep[using 1D radiation-hydro simulations by][]{2016ApJ...816...34O}.
Given that oxygen atoms are heavier than hydrogen atoms, this process of atmospheric loss is responsible for an oxygen build-up in the atmosphere \citep[e.g.][]{2015AsBio..15..119L}. 
This raises a fundamental question in astrobiology: what are appropriate biosignatures?
The observation of dioxygen or ozone was often proposed as a potential sign for life (as we know it), but atmospheric loss is a purely abiotic process which has the same signature. 

The model exposed in \citet{2017MNRAS.464.3728B} was used to estimate the water loss from the planets of TRAPPIST-1 \citep{2017A&A...599L...3B, 2017AJ....154..121B}.
In particular, TRAPPIST-1b has lost a large amount of water during the lifetime of the system (up to 100 Earth oceans). 
This appeared to be in contradiction with the estimates of masses and densities of TRAPPIST-1b done by \citet{2018A&A...613A..68G}, which favored a scenario with a large amount of volatiles. 
However, a recent work of \citet{2018ApJ...865...20D} showed a more uniform distribution of densities for the TRAPPIST-1 planets which is compatible with an increase of volatiles with distance, in agreement with our model.
As for the HZ planets, the model of \citet{2017MNRAS.464.3728B} used in \citet{2017AJ....154..121B} showed that they lost a very small amount of water before reaching the HZ.
This means that if they formed with a reasonable amount of water, the probability they still have some is high.

\subsection{Orbital and rotational evolution}

Let us concentrate first on the evolution of the eccentricity of the orbits of the TRAPPIST-1 planets.
The eccentricity is damped by the planetary tide and excited by planet-planet interactions, which leads to a state where both processes compensate each other. 
The equilibrium eccentricity determined by the competition of tidal damping and gravitational excitation is very small for the planets of TRAPPIST-1 \citep{2017NatAs...1E.129L, 2018A&A...612A..86T, 2018A&A...613A..68G}.
In particular, it is below $0.001$ for the two inner planets.

\begin{figure}[ht!]
 \centering
 \includegraphics[width=0.5\textwidth,clip]{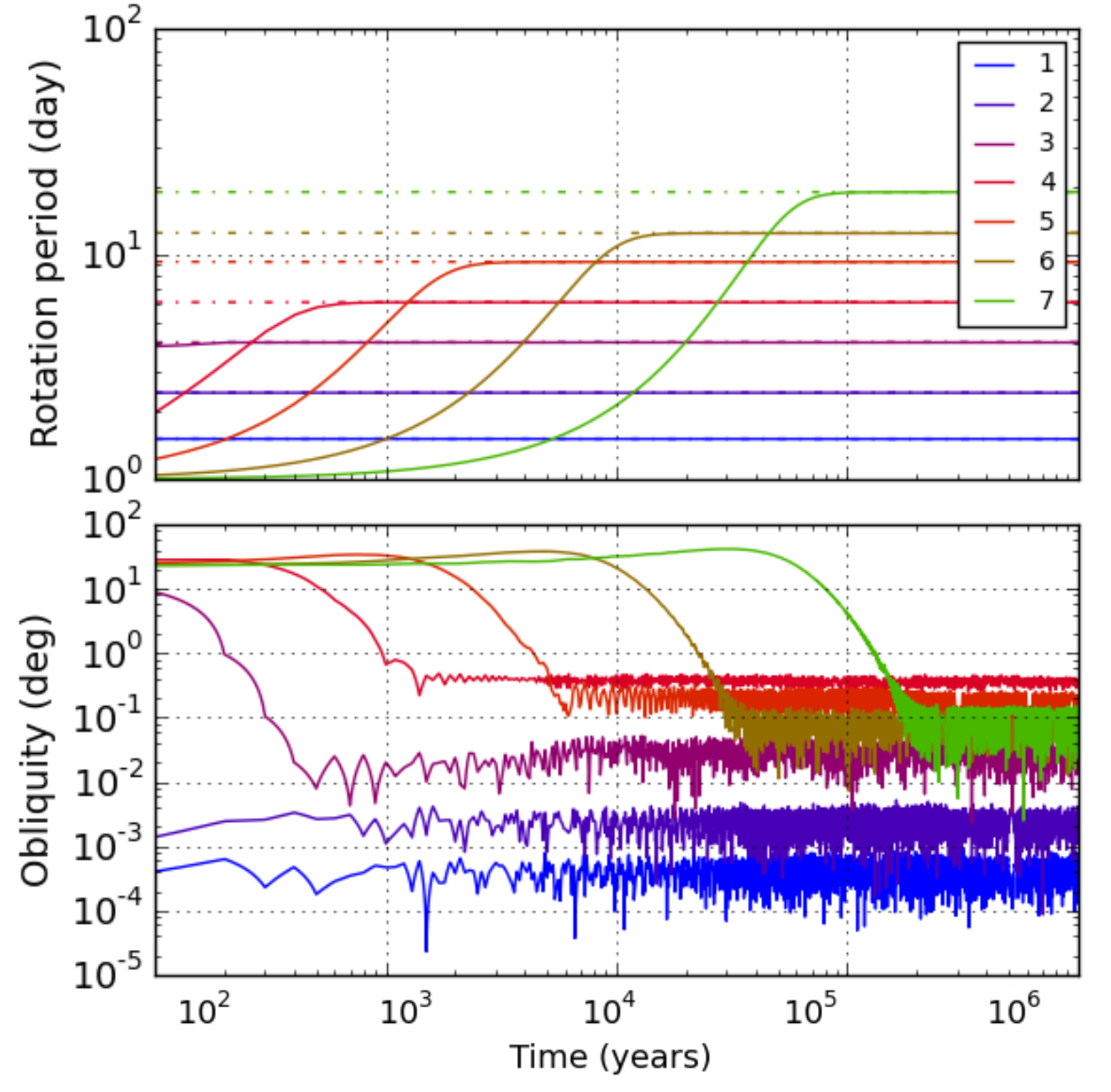}      
  \caption{Evolution of the rotation period and obliquity for the seven planets of TRAPPIST-1. On the top panel, the dash-dotted lines represent the synchronization period.}
  \label{bolmont:fig2}
\end{figure}

The rotation of the TRAPPIST-1 planets is also influenced by tides, in particular their obliquity and rotation period. 
Tides tend to damp the obliquity (and therefore reduce any seasonal variations) and synchronize the rotation (leading to a tidally locked rotation with a permanent day side and a permanent night side).
The rotation typically evolves much faster than the eccentricity. 
In the case of TRAPPIST-1 and assuming a dissipation of 1/10th of Earth's dissipation \citep{1997A&A...318..975N}, we found that the obliquity is damped and the rotation synchronized for all planets in about 200,000 years (Bolmont et al. in prep).
Figure~\ref{bolmont:fig2} shows the evolution of these quantities with time.
This means that by the age of the system \citep[a few $10^9$~yrs;][]{2017NatAs...1E.129L}, the planets should have a very small obliquity and be synchronized.

\section{Once in the HZ}

To sum up, when the planets reach the HZ, they should be on quasi-circular orbits, their rotation should be synchronized and they should have no obliquity.
Besides, the HZ planets should still have water at the age of the system.

With these additional theoretical constraints, we can use them as hypotheses to study the climate of the HZ planets using a Global Climate Model \citep{2011ApJ...733L..48W, 2011A&A...532A...1S, 2013Icar..222...81F}.
That is what was done in \citet{2018A&A...612A..86T} where the focus was on TRAPPIST-1e to h.
In agreement with the theoretical constraints, we considered water-rich worlds, tidally locked planets with no obliquity.
We also considered planets on circular orbits. 
Although it is not strictly the case here, it has been shown in \citet{2016A&A...591A.106B} that for planets around very low mass stars the difference of flux received between periastron and apoastron does not impact the climate unless the eccentricity is very high.
Thus, to study the climate of the TRAPPIST-1 planets, it is acceptable to neglect the eccentricity.
Finally, we supposed atmospheres with CO$_2$, CH$_4$, N$_2$ and H$_2$O.

\begin{figure}[ht!]
 \centering
 \includegraphics[width=0.5\textwidth,clip]{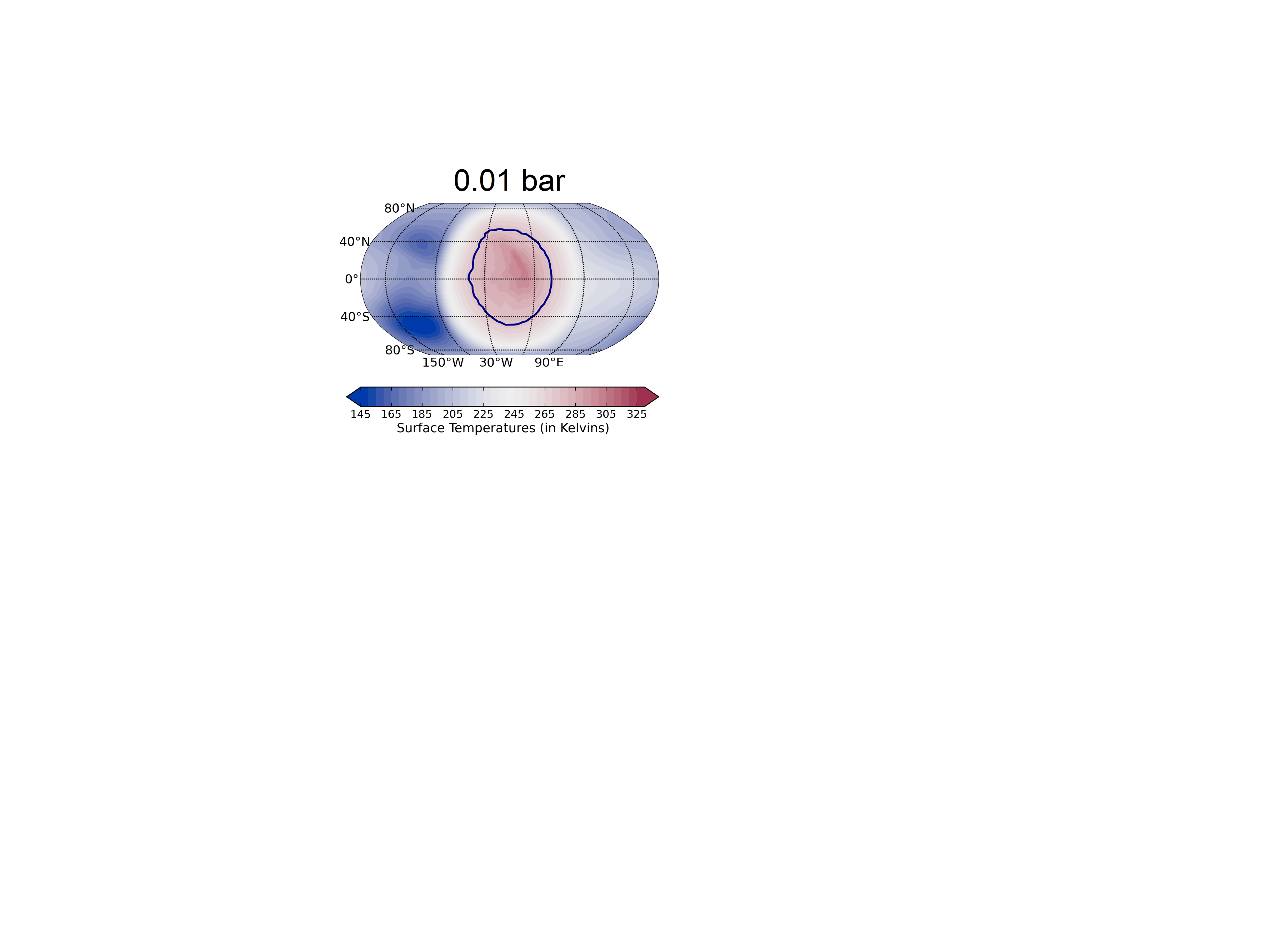}      
  \caption{Four-year average surface temperature map of TRAPPIST-1e assuming an atmosphere of 10~mbar of N$_2$ and 376~ppm of CO$_2$. The solid line contour corresponds to the delimitation between surface liquid water and sea water ice. Figure adapted from \citet{2018A&A...612A..86T}.}
  \label{bolmont:fig3}
\end{figure}

Under these hypotheses, \citet{2018A&A...612A..86T} found that TRAPPIST-1e can maintain liquid water at the substellar point for a large variety of atmospheric compositions.
Figure~\ref{bolmont:fig3} shows a temperature map of planet~e for a tenuous N$_2$ atmosphere.  
For TRAPPIST-1~f and g, we found that they can sustain surface liquid water assuming they have a few bars of CO$_2$.
However, if the CO$_2$ collapses on the night side, these planets could be trapped into a permanent snowball state.
Finally TRAPPIST-1~h is unable to maintain surface liquid water.
It cannot build up more than $10^2-10^3$~ppm of CO$_2$, whatever the amount of background gas.



\section{Discussion}
For climate study purposes, as \citet{2016A&A...591A.106B} showed, small eccentricity orbits can be considered circular. 
However, a difference in flux between periastron and apoastron is not the only effect that eccentricity can have on the climate.

In many aspects, TRAPPIST-1 is comparable to the system of Jupiter and its satellites. 
The eccentricity of Io is damped by tides and excited by the other satellites (especially by Europa and Ganymede, in mean motion resonance with Io), this leads to a small remnant equilibrium eccentricity of $\sim 0.004$.
This non-zero eccentricity leads to a tidal deformation of the satellite, which is responsible for the observed intense surface activity (tidal heat flux of $\sim 3$~W/m$^2$, \citealt{2000Sci...288.1208S}; intense volcanic acticity, \citealt{2007Sci...318..240S}).
The exact same situation is true for the planets of TRAPPIST-1. 
The tidal heat flux for each planets has been evaluated in \citet{2017NatAs...1E.129L} and \citet{2018A&A...612A..86T}.
In particular the flux of TRAPPIST-1b is always higher than Io's and the flux of planets c and d are higher than the heat flux of Earth \citep{1993RvGeo..31..267P, 2010SolE....1....5D}.
Depending on the assumption on the dissipation of the planets, TRAPPIST-1e can experience a tidal heat flux of the order of magnitude of Earth's heat flux.
The effect of this tidal heat flux on the internal structure of the planets \citep{2018A&A...613A..37B} and their climate \citep{2018A&A...612A..86T} should be investigated further (see Sylvain Breton's proceeding from this same conference).

\section{Conclusion}
TRAPPIST-1 is a laboratory for a lot of different physical processes and has an astrobiological interest enhanced by the observation prospects (with the JWST).
To assess the habitability of this system and prepare for the future atmospheric observations, it is necessary to have a comprehensive understanding of the system, from its potential insolation and dynamical history to its present dynamical state. 

\begin{acknowledgements}
E.B. thanks Martin Turbet, Franck Selsis, J\'er\'emy Leconte for fruitful collaborations on this system.
E.B. acknowledges funding by the European Research Council through ERC grant SPIRE 647383.
\end{acknowledgements}

\bibliographystyle{aa}  

\end{document}